\documentclass[print]{aa}  
\usepackage{graphicx}
\usepackage{natbib}
\usepackage{amsmath}
\usepackage{verbatim}
\usepackage[varg]{txfonts}
\begin{document}
\title{Delayed detonations in full-star models of Type Ia supernova
  explosions}

\author{F. K. R{\"o}pke\inst{1,2}
        \and
        J. C. Niemeyer\inst{3}}     
        \institute{Max-Planck-Institut f\"ur Astrophysik,
              Karl-Schwarzschild-Str. 1, D-85741 Garching, Germany
              \and
              Department of Astronomy and Astrophysics, University of
              California, Santa Cruz, CA 95064, U.S.A.
              \and
              Lehrstuhl f\"ur Astronomie, Universit\"at W\"urzburg,
              Am Hubland, D-97974 W\"urzburg, Germany
       }
   \offprints{F. K. R{\"o}pke}
\date{Received 18 October 2006 / Accepted 12 December 2006}

  \abstract
   {}
   {We present the first full-star three-dimensional explosion simulations of
   thermonuclear supernovae including parameterized
   deflagration-to-detonation transitions that occur once the flame
   enters the distributed burning regime.}
   {Treating the propagation of both the
   deflagration and the detonation waves in a common front-tracking
   approach, the detonation is prevented from crossing ash
   regions.}
   {Our criterion
   triggers the detonation wave at the outer edge of the deflagration
   flame and consequently it has to sweep around the complex structure
   and to compete with expansion. Despite the impeded detonation
   propagation, the obtained explosions show reasonable agreement with
   global quantities of observed Type Ia
   supernovae. 
   By igniting the
   flame in different numbers of kernels around the center of the
   exploding white dwarf, we set up three different models shifting
   the emphasis from the deflagration phase to the detonation
   phase. The resulting explosion energies and iron group element
   productions cover a large part of the diversity of Type Ia supernovae.}
   {Flame-driven deflagration-to-detonation
   transitions, if hypothetical, remain a possibility deserving further investigation. }

   \keywords{Stars: supernovae: general -- Hydrodynamics -- Instabilities
    -- Turbulence -- Methods: numerical}

   \maketitle
%

\section{Introduction}

The observables of Type Ia supernovae (SNe~Ia) are determined by 
the composition of the ejecta and the energy release in the 
explosion. Both depend sensitively on the density at which the 
carbon-oxygen white dwarf (WD) material is burned by the thermonuclear
flame. This in turn is a consequence of the competition between the
flame propagation and the expansion of the WD due to nuclear
energy release. Therefore, 
reproducing the characteristics of SNe~Ia in a
thermonuclear white dwarf explosion model requires a specific sequence
of flame velocities. 

After the thermonuclear flame has ignited near the
center of the WD, two distinct modes of outward
propagation are hydrodynamically admissible. In a subsonic
deflagration the flame is mediated by the thermal conduction of the
degenerate electrons, while a supersonic detonation is driven by
shock waves. Observational constraints rule out a prompt detonation
\citep{arnett1969a} and thus
the flame has to start out as a deflagration \citep{nomoto1976a}. Such a
scenario leads to a stratification of dense fuel
atop light ashes. Due to Rayleigh-Taylor and
Kelvin-Helmholtz instabilities, turbulence is generated. This results
in a turbulent cascade with eddies decaying into ever smaller ones
until they are dissipated at the Kolmogorov scale ($\lesssim 1 \,
\mathrm{mm}$). In the cascade, the turbulent velocity decreases with
length scale. Thus, turbulent motions interact with the flame down to
the so-called Gibson length $l_\mathrm{G}$, where the turbulent
velocity fluctuations $v'$ become comparable to the laminar flame
speed $s_\mathrm{l}$,
\begin{equation*}
v'(l_{\mathrm{G}}) \approx s_\mathrm{l}.
\end{equation*}
For most parts of the explosion, $l_{\mathrm{G}}$ is much larger than
the flame width $l_{\mathrm{f}}$, indicating that the internal flame structure is
unaffected by turbulence. Hence the interaction in this \emph{flamelet
  regime} of turbulent combustion is purely kinematic, corrugating the
flame and accelerating it by increasing its surface.
However, as the WD star expands, the fuel density drops,
$s_{\mathrm{l}}$ decreases, and $l_{\mathrm{f}}$ increases \citep{timmes1992a}. As soon as 
$l_{\mathrm{G}} \lesssim l_{\mathrm{f}}$, turbulent eddies distort the
internal flame structure. This \emph{distributed burning
  regime} commences when the fuel density has dropped close to values at
which deflagration burning ceases \citep{niemeyer1997b, niemeyer1997d, roepke2005a}.

In order to burn significant fractions of the WD, a high flame
propagation speed is necessary. Even a turbulence-boosted flame is
relatively slow so that it may not lead to a sufficient fuel
consumption before fuel densities drop below the burning threshold
($\lesssim$$10^7 \, \mathrm{g}\, \mathrm{cm}^{-3}$). Although turbulent
deflagration models
\citep{reinecke2002d,gamezo2003a,roepke2005b,roepke2006a,schmidt2006a}
reproduce the observational features of weak SN Ia
explosions \citep{blinnikov2006a}, there exist open questions \citep{kozma2005a}
and presently it seems that they cannot account
for the more energetic events. This issue certainly needs
more exploration (in particular the buoyancy induced turbulence
effects at the high Reynolds numbers of $\sim$$10^{14}$ expected in
  SNe~Ia, \citealp[see][]{cabot2006a}) but it may also be interpreted as an
incompleteness of such models. If more burning is necessary than the
deflagration model
allows for, a deflagration-to-detonation transition (DDT) would
provide the ultimate flame acceleration. The problem of such
\emph{delayed detonation models} \citep{khokhlov1991a,woosley1994a} is
the absence of a robust physical mechanism that allows predicting a
DDT in thermonuclear supernova
explosions \citep{niemeyer1999a,roepke2004a,roepke2004b}. Therefore
invoking a DDT makes it necessary to introduce a free parameter into the
model.

Examples of such parametrized models were
recently given. \citet{gamezo2005a} assume the DDT to take place at an
arbitrarily chosen instant and location, while \citet{golombek2005a}
impose a DDT once the flame enters the distributed burning regime.

Here we present the first three dimensional simulations of delayed detonations
based on the numerical methods proposed by
\citet{golombek2005a} and adopting their DDT parametrization. Since
the only instant where the flame properties 
change drastically is the beginning of the distributed burning regime,
it seems natural to assume a DDT here \citep{niemeyer1997b}.
Although presently there is no indication of a DDT occurring in the distributed
burning regime
\citep{lisewski2000b,zingale2005a}, more exploration is necessary to
settle this question.
Our DDT parametrization does not \emph{a priori} fix the instant
of its occurrence nor its location on the flame and thus a failure
triggering a viable detonation is possible.
If successful, however, the detonation propagation is likely to proceed
asymmetrically. In order to avoid artificial symmetries in the setup of the
simulations, we pursue a full-star implementation of the model.  
Our aim is to determine whether such a model is consistent with global properties of
observed SNe~Ia, which is non-trivial since here the DDT is not used
to tune the model but results from a physically motivated hypothesis. 

\section{Modeling  approach}

Aiming at parameterizing the location and time of the DDT according to
properties of the deflagration flame, a self-consistent treatment of
the interaction of the flame with turbulence is essential. This is
achieved by a large-eddy simulation approach on the basis of methods
presented by
\citet{niemeyer1995b}, \citet{reinecke1999a}, \citet{reinecke2002b}, \citet{roepke2005c}, and
\citet{schmidt2006c}. In particular, the deflagration flame is
represented as a sharp interface between fuel and ashes applying the
level set method and unresolved turbulence is described via
a subgrid-scale model from which the turbulent flame propagation speed is determined. 

\cite{maier2006a} found that even tiny regions of ash stall the
detonation wave. Consequently, the detonation has to
sweep around the complex deflagration structure. This requires a
versatile implementation of the detonation. Following the suggestion of
\citet{golombek2005a} we model the detonation wave in the same
approach as the deflagration flame applying a second level set
treatment (see this reference for thorough tests of the
implementation). 

Lacking a sound theoretical understanding of the mechanism
of the DDT, we restrict our model to a single DDT which is
achieved by initiating the detonation level set the first patch of the
deflagration flame where $l_{\mathrm{f}} \ge 0.98\,
l_{\mathrm{G}}$. Although this approach may seem rather
artificial, it was chosen to simplify matters and to obtain a clear
picture of the detonation propagation through the WD. Physically,
it is motivated by the assumption that a DDT is a rare event.

In fuel regions the propagation speed of the detonation
front is prescribed according to \citet{sharpe1999a}, while
propagation into ash is suppressed.
One question to be addressed in the present study is to which degree
such a handicapped detonation is capable of burning major parts of the
WD matter -- an issue calling for three-dimensional simulations. 

Our description of the nuclear reactions is based on \citet{reinecke2002b}
with the exception that the lower thresholds are set
differently for burning C and O, following
\citet{khokhlov1991a}. While we allow C consumption down 
to fuel densities of $10^6 \, \mathrm{g} \, \mathrm{cm}^{-3}$, O
consumption ceases already at $10^7 \, \mathrm{g} \, \mathrm{cm}^{-3}$.
The WD is constructed using a suitable equation of state assuming
isothermality at $T = 5 \times 10^5 \, \mathrm{K}$, a central
density of $2.9 \times 10^9 \, \mathrm{g} \, \mathrm{cm}^{-3}$, and a
composition of equal parts by mass of C and O. It is set up on a
Cartesian computational grid of $256^3$ cells with outflow boundary
conditions on all sides. As described by
\citet{roepke2006a}, we follow the propagation of the deflagration
flame and the WD expansion with two nested moving grid patches optimizing
the flame resolution.

\section{Simulations}

\begin{figure}[t]
\centerline{\includegraphics[width = \linewidth,viewport = 0 0 340 717,clip]{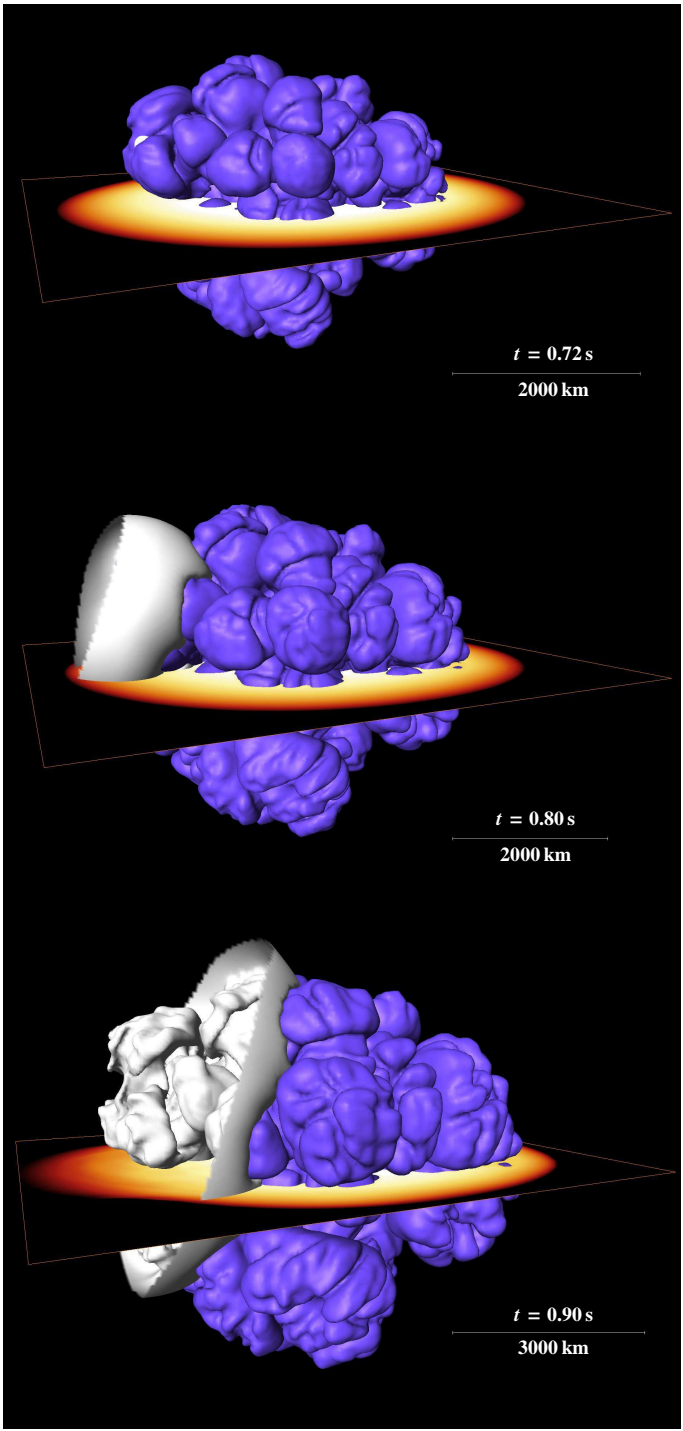}}
\caption{Initiation and propagation of the deflagration wave (white
  isosurface) in Model DD\_020. The deflagration flame is shown as
  dark isosurface and the extend of the star is indicated by the
  central plane mapping the logarithm of the density.\label{fig:evo_dd}}
\end{figure}

With the DDT parametrization fixed as above, the only major undetermined
parameter that impacts the evolution of the model is the configuration
of the deflagration flame ignition. Although not attempting a thorough
exploration of the parameter space here, we set up three different
models in the multi-spot scenario similar to the simulations presented
by \citet{roepke2006a}. This allows us to test the robustness of the
imposed DDT criterion. By placing 5 (Model DD\_005), 20 (Model DD\_020),
and 800 (Model DD\_800)  flame ignition kernels inside a radius of
$180 \, \mathrm{km}$ around the WD's center, the efficiency of burning
in the deflagration phase is significantly altered. While Model
DD\_005 leads to relatively little burning, igniting 800
flames results in a strong deflagration phase. We focus our
description on the intermediate case of Model DD\_020. 

After ignition,
the deflagration phase proceeds in a similar way as described by
\citet{roepke2006a}. The burning bubbles grow by flame propagation and
rise buoyantly towards the surface of the WD. Due to instabilities and
partial merger of the bubbles, a complex connected structure
develops. The parametrized DDT criterion is met first $0.724 \,
\mathrm{s}$ after ignition at the outer edge of the flame front. Here,
the detonation is triggered by initiating the corresponding level set,
as shown in the upper panel
of Fig.~\ref{fig:evo_dd}. The fact that the outer edge of the
deflagration flame is chosen by our DDT criterion is not
surprising. Turbulence is generated preferentially at large buoyant
bubbles and at the same time the density is lowest at the outermost
parts of the flame making the flame broadest here. This favors the
transition to the distributed burning regime and thus the parametrized
initiation of the detonation wave.

The center and lower panels of Fig.~\ref{fig:evo_dd} show the
propagation of the detonation wave. As implemented in our model, it
cannot cross ash regions and, burning towards the star's center, it
therefore wraps around the corrugated
deflagration structure. In this way, it takes about $0.2 \,
\mathrm{s}$ before it arrives at the center of the WD. Meanwhile, the
star keeps expanding and the deflagration continues in regions not yet
reached by the detonation wave. Consequently, the density
of the fuel ahead of the detonation drops quickly after passing the WD's
center. In Model DD\_020 burning stalls shortly before the detonation
reaches the far side of the deflagration structure. 
This still implies burning of most of the WD,
since these deflagration features have
already reached the low density edge of the star. 

The DDT parameters and the results of all three simulations
simulations are summarized in Table~\ref{tab:res}. The more
kernels the deflagration is ignited in, the less energetic becomes the
explosion and the less NSE elements are synthesized. The latter effect is
partially compensated by more IME produced in weaker explosions giving
rise to comparable masses of total burned material. In terms of the masses of the
explosion products, the entire fuel remaining from the deflagration is
affected by the detonation in all simulations. This is the reason for
the virtually complete C-depletion
in all models. The different expansion history, however, is reflected
by up to $\sim 0.2\, M_{\odot}$ of O originating from material
detonating at densities less than $10^7 \, \mathrm{g} \,
\mathrm{cm}^{-3}$ .

Although the fuel density at the location of the DDT increases with
the number of deflagration ignition spots and the DDT sets in earlier,
the explosions do not become stronger. The most important parameter
for the overall explosion strength is the nuclear energy release in
the deflagration phase prior to the DDT,
$E_{\mathrm{nuc}}^{\mathrm{DDT}}$. It accounts for the expansion 
of the star and therefore determines which fraction of fuel is
available for the detonation to burn to NSE or IME. 
Obviously, the asymptotic kinetic energy is larger in models releasing
less nuclear energy prior the the DDT (cf.\ Table~\ref{tab:res}). 
Moreover, the complexity
of the deflagration flame increases with more ignition spots impeding
and delaying the propagation of the detonation. These effects cause a
larger number of ignition spots to shift the emphasis of the explosion model
from the detonation phase to the deflagration phase which is
characterized by the ratio of the mass of nuclear ash produced by the
deflagration to that produced by the detonation,
$\mathcal{R}^{\mathrm{def}}_{\mathrm{det}}$ (see
Table~\ref{tab:res}). Whereas in Model DD\_800 the
deflagration mass consumption exceeds that of the detonation by a
factor of 1.59, the detonation burns
about three times as much material as the deflagration in Model DD\_005.

This way, we obtain three different final configurations.
Assuming that a large part of the NSE material is actually $^{56}$Ni,
which by radioactive decay powers the light curve of the event,
Model DD\_005 would correspond to a very bright supernova. Model DD\_020 provides an
intermediate case with values similar to the one-dimensional W7 model
\citep{nomoto1984a}, which is considered to be prototypical for many
SNe~Ia, while Model
DD\_800 corresponds to a dimmer event. In the latter, the detonation does
not significantly enhance the NSE production over similar pure
deflagration models \citep{roepke2006a}.  The major difference to pure
deflagration models is a more complete burning of the outer parts of
the WD leading to larger asymptotic kinetic
energies of the ejecta and a different composition of these. The inner
core is dominated by NSE material, partially mixed with IME
originating from the deflagration phase and low-density detonation
burning. Above, a layer dominated by IME is found which is completely
C-depleted in agreement with a recent analysis of observed spectra
\citep{stehle2005a}. 

The asymptotic kinetic energies of the ejecta range
around the expected value of $\sim$$10^{51}\, \mathrm{erg}$.
In all tested scenarios, the DDT occurs before the
deflagration structure encloses large pockets of fuel, which would be
unreachable for the detonation wave. This, however, may in part be due to
the low resolution of the simulations and needs further exploration.

\begin{table*}
\centering
\caption{Model parameters: the DDT columns provide values for the time
  of the initiation of the detonation wave ($t_{\mathrm{DDT}}$), for
  its distance from the center of the WD, for the density at the DDT spot
  ($\rho_{\mathrm{DDT}}$), and for the nuclear energy release
  in the deflagration prior to the DDT
  ($E_{\mathrm{nuc}}^{\mathrm{DDT}}$). The last six columns
  characterize the asymptotic kinetic energy of the ejecta
  ($E_{\mathrm{kin}}^{\mathrm{asympt}}$) and the composition of the
  final stages of the explosion models, and give the ratio
  $\mathcal{R}^{\mathrm{def}}_{\mathrm{det}}$ of the
  masses of nuclear ash produced by the deflagration to that produced
  by the detonation.
\label{tab:res}}
\begin{tabular}{rr|rrrr|rrrrrr}
\hline\hline
\multicolumn{1}{p{0.038 \linewidth}}{model} &
\multicolumn{1}{p{0.093 \linewidth}}{deflagration} &
\multicolumn{4}{|c|}{{DDT}} &
\multicolumn{6}{|c}{final stage ($t = 10 \, \mathrm{s}$)}\\
&\multicolumn{1}{p{0.093 \linewidth}}{ignition spots}&
\multicolumn{1}{|p{0.03 \linewidth}}{$t_{\mathrm{DDT}}$ [s]}&
\multicolumn{1}{p{0.105\linewidth}}{distance from center [$10^8 \, \mathrm{cm}$]}&
\multicolumn{1}{p{0.08 \linewidth}}{$\rho_{\mathrm{DDT}}$ [$ 10^7 \,\mathrm{g} \,
  \mathrm{cm}^{-3}$]}&
\multicolumn{1}{p{0.04 \linewidth}}{$E_{\mathrm{nuc}}^{\mathrm{DDT}}$ [foe]}&
\multicolumn{1}{|p{0.05 \linewidth}}{$E_{\mathrm{kin}}^{\mathrm{asympt}}$ [foe]}&
\multicolumn{1}{p{0.06 \linewidth}}{$M(\mathrm{NSE})$ [$M_\odot$]}&
\multicolumn{1}{p{0.058 \linewidth}}{$M(\mathrm{IME})$ [$M_\odot$]}&
\multicolumn{1}{p{0.058 \linewidth}}{$M(\mathrm{C})$ [$M_\odot$]}&
\multicolumn{1}{p{0.052 \linewidth}}{$M(\mathrm{O})$ [$M_\odot$]}&
\multicolumn{1}{p{0.03 \linewidth}}{$\mathcal{R}^{\mathrm{def}}_{\mathrm{det}}$}\\
\hline
DD\_005 & 5   & 0.731 & 1.733 & 1.33 & 0.221 & 1.524 & 1.141 &
0.220 & 0.00586 & 0.0386 & 0.37\\
DD\_020 & 20  & 0.724 & 1.733 & 1.92 & 0.500 & 1.237 & 0.833 &
0.435& 0.0190 & 0.118 & 1.48\\
DD\_800 & 800 & 0.675 & 1.820 & 2.40 & 0.656 & 1.004 & 0.638 & 0.547 &
0.0297 & 0.189 & 1.59\\
\hline
\end{tabular}
\end{table*}

\section{Conclusions}

In the presented study, we imposed a hypothetical DDT criterion on SN
Ia explosion simulations. Being guided by a physical scenario (the
transition of the deflagration flame from the flamelet to the
distributed burning regime), it cannot be used to tune the
models. Thus, the success of the
presented simulations to reproduce gross features of
SNe~Ia, although it cannot be taken as a confirmation of the
hypothesis, leaves the possibility of a flame-driven DDT open to a
more detailed investigation of potential physical mechanisms.
This result is not obvious, since the time and location of the DDT is not
fixed and it thus may occur at instances disfavoring an efficient
detonation. However, the three simulations presented here trigger a
successful detonation after very different deflagration phases
indicating robustness of the criterion. 
Parameterizing the DDT according to properties of the turbulent
deflagration flame is naturally very sensitive to the turbulence
description and we emphasize the importance of a well-founded
subgrid-scale model.

The three-dimensional simulations presented here produce somewhat less
energetic explosions than the two-dimensional setups of
\citet{golombek2005a}. This is most likely due to the symmetry
constraints in the two-dimensional simulations where the detonation
wave is initiated in two toroidal rings. Moreover, deflagrations
are more energetic in three dimensions and produce a more complex
structure do to the additional degree of freedom, both effects
hampering the detonation wave propagation. Nonetheless, almost all
WD material is subject to nuclear burning in the models studied here.  
Yet the detonation wave does not in all cases outrun the  
deflagration features on the far side of
the DDT. This leads to asymmetries in the structure of the outermost
ejecta which, however, contain little mass. Such asymmetry effects may
be an artifact of our restriction to a single DDT. The
DDT criterion can easily be modified to account for
multiple DDTs at each patch of the flame where the distributed burning
regime is entered. Such parameters of the DDT criterion will certainly
have an impact on the explosion which will be analyzed in a
forthcoming study. Although we expect that the range of
  variation in the results will be similar to that found with a single
  DDT, the correlation between the number of ignition kernels and the
  explosion strength will possibly shift towards stronger explosions
  for denser ignitions.

The results of the explosion simulations are in good agreement with
current observational constraints of SNe~Ia.
Our set of models indicates that the presented delayed detonation
scenario can cover at least large parts of the observed
sample of SNe~Ia with respect to explosion energies and
NSE material production. 
However, for a thorough comparison with observational results, a
detailed analysis of the ejecta composition on the basis of
high-resolution models is required and will be addressed in a
forthcoming study. 

\begin{acknowledgements}
We thank Irina Golombek for providing her implementation of the
DDT and Wolfgang Hillebrandt and Paolo Mazzali for helpful
discussions. The research of
JCN was supported by the Alfried Krupp Prize for Young 
University Teachers of the Alfried Krupp von Bohlen und Halbach
Foundation.
\end{acknowledgements}

\end{document}